# Deep learning analysis of intracranial EEG for recognizing drug effects and mechanisms of action

Konstantin Y. Kalitin [a,b*1], Alexey A. Nevzorov [c2], Denis A. Babkov[a,b3], Alexander A. Spasov [a,b4], Olga Y. Mukha [a5]

[a] *Department of Pharmacology and Bioinformatics, Volgograd State Medical University, Address: 1, Pavshikh Bortsov Sq., Volgograd 400131, Russian Federation*

[b] *Research Institute of Pharmacology, Address: Ulitsa Novorossiyskaya, 39, Volgograd 400087, Russian Federation*

[c] *LIT RAS - Branch of FSRC "Crystallography and Photonics" RAS, Address: 1, Svyatoozerskaya str. , Shatura, 140700, Russian Federation*

[*]Corresponding author

Konstantin Y. Kalitin; Email address: ky.kalitin@gmail.com; Address: 1, Pavshikh Bortsov Sq., Volgograd 400131, Russian Federation; Telephone: +7(8442)942423

Abbreviations[6]

---

**Abstract**

Predicting novel drug properties is fundamental to polypharmacology, repositioning, and the study of biologically active substances during the preclinical phase. The use of machine learning, including deep learning methods, for the identification of drug-target interactions has gained increasing popularity in recent years.

[1] Email address: ky.kalitin@gmail.com
[2] Email address: terrapevt@mail.ru
[3] Email address: dababkov@volgmed.ru
[4] Email address: aaspasov@volgmed.ru
[5] Email address: olay.myha14@gmail.com

[6] DTI, drug-target interaction; EEG electroencephalogram; GABA, gamma-aminobutyric acid; MEA microelectrode arrays; OB, olfactory bulbs; CNS, central nervous system.

The objective of this study was to develop a method for recognizing psychotropic effects and drug mechanisms of action (drug-target interactions) based on an analysis of the bioelectrical activity of the brain using artificial intelligence technologies.

Intracranial electroencephalographic (EEG) signals from rats were recorded (4 channels at a sampling frequency of 500 Hz) after the administration of psychotropic drugs (gabapentin, diazepam, carbamazepine, pregabalin, eslicarbazepine, phenazepam, arecoline, pentylenetetrazole, picrotoxin, pilocarpine, chloral hydrate). The signals were divided into 2-second epochs, then converted into 2000x4 images and input into an autoencoder. The output of the bottleneck layer was subjected to classification and clustering using t-SNE, and then the distances between resulting clusters were calculated. The neural network model was validated through 5-fold cross-validation.

The classification accuracy obtained for 11 drugs during cross-validation was 0.58 ± 0.013. The resulting classification accuracy was statistically higher than the baseline accuracy of a random classifier, which was 0.0909 ± 0.045 ($p<0.0001$). t-SNE maps were generated from the bottleneck parameters of intracranial EEG signals. The relative proximity of the signal clusters in the parametric space was assessed.

The present study introduces an original method for biopotential-mediated prediction of effects and mechanism of action (drug-target interaction). This method employs convolutional neural networks in conjunction with a modified selective parameter reduction algorithm. Post-treatment EEGs were compressed into a unified parameter space. Using a neural network classifier and clustering, we were able to recognize the patterns of neuronal response to the administration of various psychotropic drugs.

## 1 Introduction

Drug-target interaction (DTI) prediction has become one of the most significant techniques in pharmacology-related areas, including polypharmacology, drug repositioning, and drug discovery, as well as side-effect and drug resistance prediction [1]. At present, the ligand-based, target-based, and chemogenomic methods are recognized as three key computational approaches for predicting DTI [2].

The ligand-based approach predicts interactions by comparing the chemical structure of a new ligand to known ligands. However, ligand-based methods are ineffective if there is an insufficient number of known ligands [3].

For target-based approaches (e.g., docking simulation methods), the target protein structures are necessary for simulation, which becomes inapplicable when pharmacological targets are unknown or the structures of target proteins are too complex to obtain [3,4].

A chemogenomic approach has successfully been applied in drug discovery and repositioning. This method integrates the chemical space of ligands and the genomic space of target proteins into a unified pharmacological space. In the case of a chemogenomic approach for psychotropic drugs, the major challenge is the scarcity of known ligands, targets, and drug-protein interactions [5,6].

The specificity of psychotropic drug action is reflected in the bioelectric activity of the brain [7]. By binding to targets with complementary structures, psychotropic drug molecules modify the electrical behavior of neurons and lead to specific electroencephalogram (EEG) reactions [8].

These biological reactions correlate with target binding and can be analyzed using computational models. This concept is supported by abundant evidence from several previous pharmaco-EEG studies. Pharmaco-EEG is very important in the development of central nervous system (CNS)-active compounds, as well as drug resistance and side-effect prediction [9]. Various validated methods are available for exploring the mechanism of action and effects of drugs on the brain [10]. The modulation of serotonergic, dopaminergic, noradrenergic, cholinergic, or opioidergic neurotransmission causes specific changes in EEG frequency [11,12,13]. Different drugs may induce the same EEG pattern, and despite the large interindividual variability, many EEG features are considered characteristic and unique to specific compounds [8,14,15].

Machine learning methods help extract information from EEG recordings, and with increasing frequency, are considered perspective neuroscientific tools [16,17]. Convolutional neural network (CNN) is currently the dominant deep learning strategy for EEG classification [18]. CNNs are widely adopted to features of EEG signals to assess the magnitude of pharmacological effects; it can specifically be used to predict and monitor the depth of anesthesia [19,20,21].

DTI prediction methods use both biological and chemical features of ligands and targets, and different machine learning techniques can be employed [3,5,22].

Here, we propose a new intracranial EEG (i-EEG)-mediated deep learning method that addresses DTI prediction as an i-EEG classification and clustering task in multidimensional space (bottleneck feature space between the encoder and the decoder), which represents similarities between the mechanisms of action of drugs and their effects on the CNS (Figure 1). This approach is based on the assumption that drugs with similar effects on EEG are likely to interact with similar pharmacological targets and elicit similar psychotropic effects. It allows inferring potentially new mechanisms and effects of new compounds under study that are mapped close to known drugs in the unified parameter space.

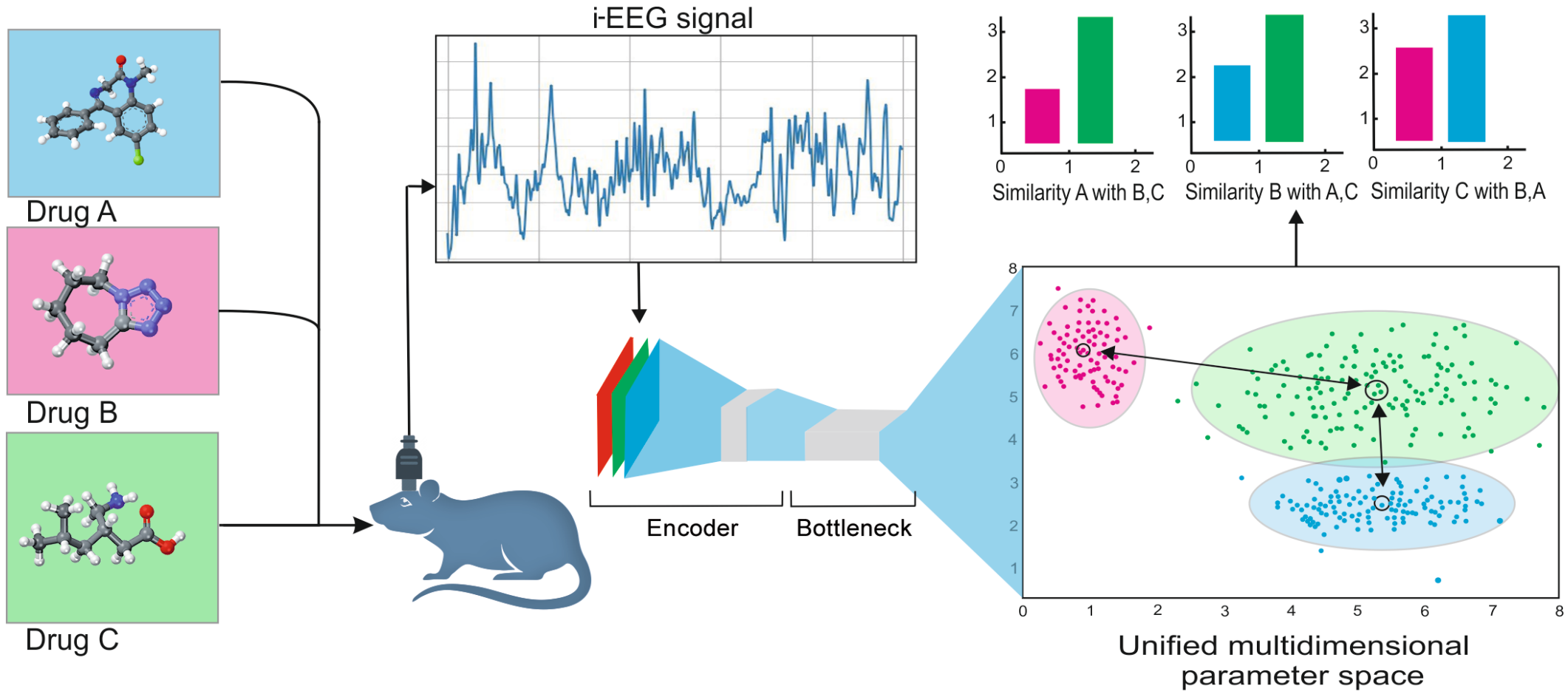

**Figure 1.** Overall study design. Intracranial electroencephalogram (i-EEG) is recorded from the animal after drug administration and fed into the neural network. The outputs of the bottleneck layer are clustered and then the distances between the clusters are calculated.

## 2 Materials and methods

### 2.1 Subjects

The experiments were carried out using inbred male rats weighing 390–500 g obtained from the Rappolovo Animal Breeding Center (Leningrad Oblast, Russian Federation). The animals were kept under a 12/12-h light/dark cycle and standard vivarium conditions (air temperature 23 ± 2 °C, continuous air supply and exhaust ventilation, and sanitary control) with free access to water and food.

### 2.2 Electrodes

The electrodes are made of silver/silver chloride (Ag-AgCl) wire (260 μm, RK 50-7-22) (Figure 2A) and are connected to a socket-contact (BL-T; resistance less than 0.01 Ω) with an insulated copper cable (MGTF 0.03; length 20–30 mm) (Figure 2B). The electrode is coated with two layers of heat-shrinkable tubing (polyethylene terephthalate); the first yellow layer (Figure 2C) serves as an insulator, whereas the second blue layer (Figure 2D) restricts electrode insertion. The electrodes were then assembled into an 8-channel connector (BLD-8, glass-filled nylon) (Figure 2E). Each electrode assembly was assigned an identification number.

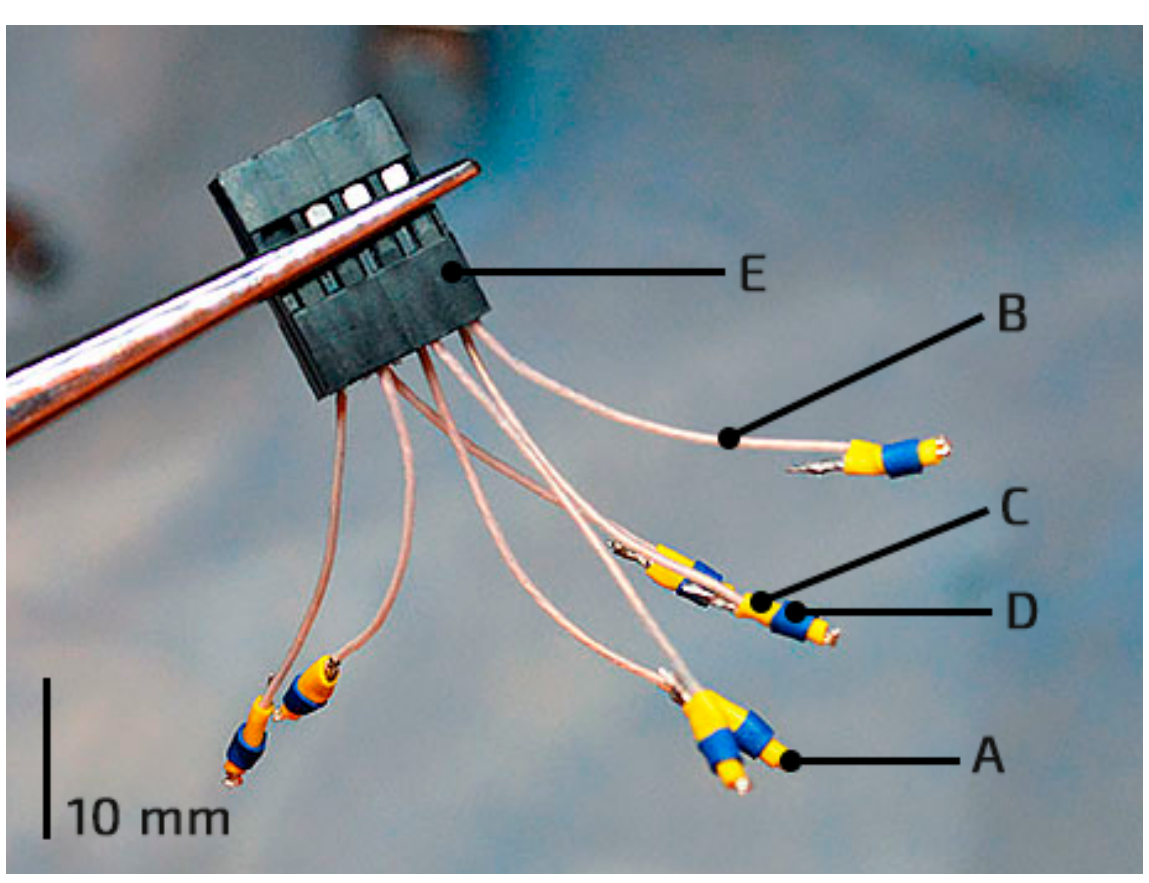

**Figure 2.** Electrode assembly for intracranial electroencephalogram (i-EEG) recordings in rats. (A) electrode; (B) copper cable; (C) insulator; (D) restrictor; and (E) connector.

**2.3 Electrode implantation**

Implantation was performed under general anesthesia with urethane at a dose of 1.5 mg/kg. The animals were stabilized using stereotaxic instruments (SR-5R-HT; Narishige Group, Tokyo, Japan) with blunt ear bars. Trepanation holes (1 mm diameter) were drilled into the skull at the designated points (Table 1, Figure 3) with periodic interruptions to avoid bone overheating.

**Table 1.** Stereotactic coordinates of electrodes and fixing screws

| Rostrocaudal localization | Left | Right |
| --- | --- | --- |
| Olfactory bulbs | FS (6.60; 2.00) | Ground OB (6.60; -2.00) |
| Primary motor cortex | A1 (0.00; 2, 00) | A2 (0.00; -2.00) |
| Primary somatosensory cortex | P3 (-4.08; 2.00) | P4 (-4.08; -2.00) |

| Secondary visual cortex | O1 (-7.08; 2.00) | O2 (-7.08; -2.00) |
|---|---|---|
| Cerebellum | FS (0; -11.64) | |

FS, fixing screw; OB, olfactory bulbs.

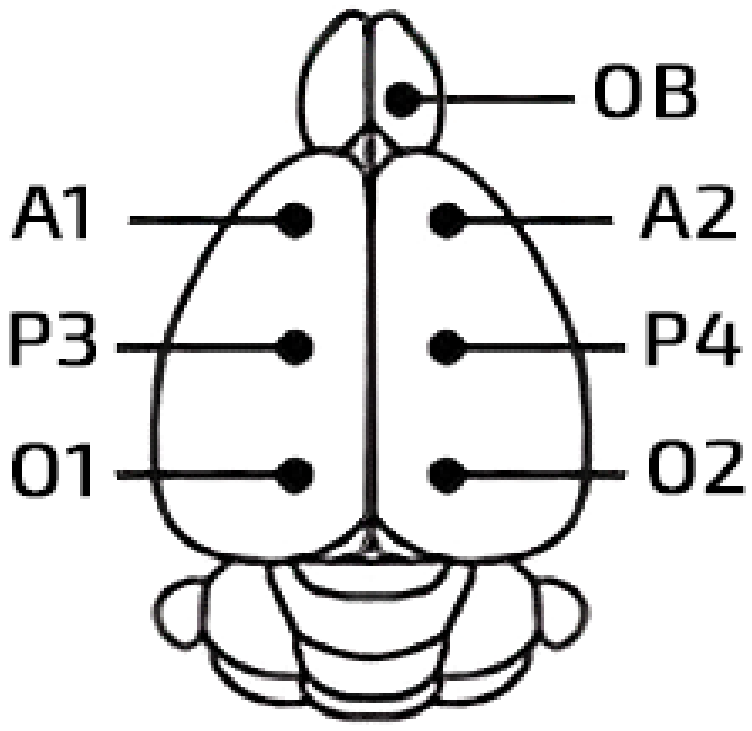

**Figure 3**. Schematic representation of the electrode placement for intracranial electroencephalogram (i-EEG) recording in a rat.

Two stainless-steel fixing anchor screws were mounted into the skull. The electrodes were inserted after epidural administration in trepanation holes and fixed with self-hardening polymer (Protacryl-M; Stoma, Kharkiv, Ukraine). The connector was glued directly to the skull and fixing screws. The connector was closed with a plastic cap to protect it from contamination and mechanical damage. All rats were then transferred to a temperature-controlled recovery environment to reach a normal state before recording commenced.

**2.4 Experimental protocol**

After 7 days of rehabilitation, the animals were randomized and administered various anticonvulsants at the maximum single therapeutic dose (conversion factor from humans to rats used in this study was 5.9 as recommended [23]) or subconvulsive dose for substances with proconvulsant activity (Table 2). After reaching the peak concentration (depending on the pharmacokinetic properties of the drug) under control of the operator, brain activity was recorded for 10 min.

**Table 2**. Characteristics of psychotropic drugs used in this study

<table>
<tr><th>Substance</th><th>Mechanism</th><th>Effect</th></tr>
<tr><td>Diazepam (6 mg/kg, po)<br>Phenazepam (1 mg/kg, po)<br>Chloral hydrate (100 mg/kg, ip)</td><td>GABA-mimetics</td><td rowspan="3">Anticonvulsant</td></tr>
<tr><td>Pregabalin (60 mg/kg, po)<br>Gabapentin (360 mg/kg, po)</td><td>Calcium channel inhibitors</td></tr>
<tr><td>Carbamazepine (200 mg/kg, po)<br>Eslicarbazepine (160 mg/kg, po)</td><td>Sodium channel inhibitors</td></tr>
<tr><td>Corazol (pentylenetetrazole; 20 mg/kg, ip)<br>Picrotoxin (2 mg/kg, ip)</td><td>GABA antagonists</td><td rowspan="2">Proconvulsant</td></tr>
<tr><td>Pilocarpine (60 mg/kg, ip)<br>Arecoline (40 mg/kg, ip)</td><td>Cholinomimetics</td></tr>
</table>

GABA, gamma-aminobutyric acid; ip, intraperitoneal injection; po, administration *per os*.

### 2.5 i-EEG Recording

A laboratory electroencephalograph (NVX-36; MKS, Moscow, Russian Federation) was used to record bioelectrical activity. Intracranial EEG signals were recorded at a sampling rate of 500 Hz, in a bipolar montage. Electrode impedance was < 5 kΩ.

i-EEG montage (Table 1): Olfactory bulbs (OB) (ground); P3-A1 (channel 1); O1-A1 (channel 2); P4-A2 (channel 3); O2-A2 (channel 4).

All data were uploaded to the cloud and logged in a spreadsheet. The EEG signals used in the dataset are available online (http://dx.doi.org/10.17632/gmkbhj28jh.1).

### 2.6 Dataset structure

Each subject had some individual characteristics that affected both the background EEG signal and changes in the EEG signal associated with drug exposure. Therefore, the dataset was compiled in a way to increase the diversity of individually associated signal features, which justifies the use of inbred animals of different ages and weights. Signals during drug exposure were recorded from five different rats for each drug.

The total size of the dataset for algorithm learning included 16,500 samples. Each sample had a duration of 2 s and a sampling rate of 500 Hz. As the EEG was recorded simultaneously from four derivations, one signal in the dataset contained 4000 data points. Additional filters were not used because they do not significantly affect CNN training and result accuracy [18].

In 5-fold cross-validation, the whole dataset (including EEG samples from all individuals for all drugs) was divided into five equally sized folds using four for training and one for prediction. Each fold was used exactly once for prediction. Resulting statistical metrics were averaged over all five runs.

### 2.7 Hardware and software specifications

The deep learning system was trained on a workstation equipped with one GPU GeForce GTX 660 (Nvidia, Santa Clara, CA). The neural network model was built and trained using the Tensorflow Python library. Clustered data was visualized using the t-distributed stochastic neighbor embedding (t-SNE) method [24].

## 3 Calculation

EEG signals cannot be modeled by some known parametric function of time; however, they can be approximated with any accuracy by an artificial neural network. In addition, neural networks are suitable for solving the problem of EEG signal classification with a rich spectrum that has no analytic representation [25].

Considering the efficiency of the neural network as a classifier and parameterizer of complex EEG signals, a unique architecture was proposed (Figure 4). The CNN architecture was built taking into account that networks for processing quasi-periodic signals do not require much depth [26,27].

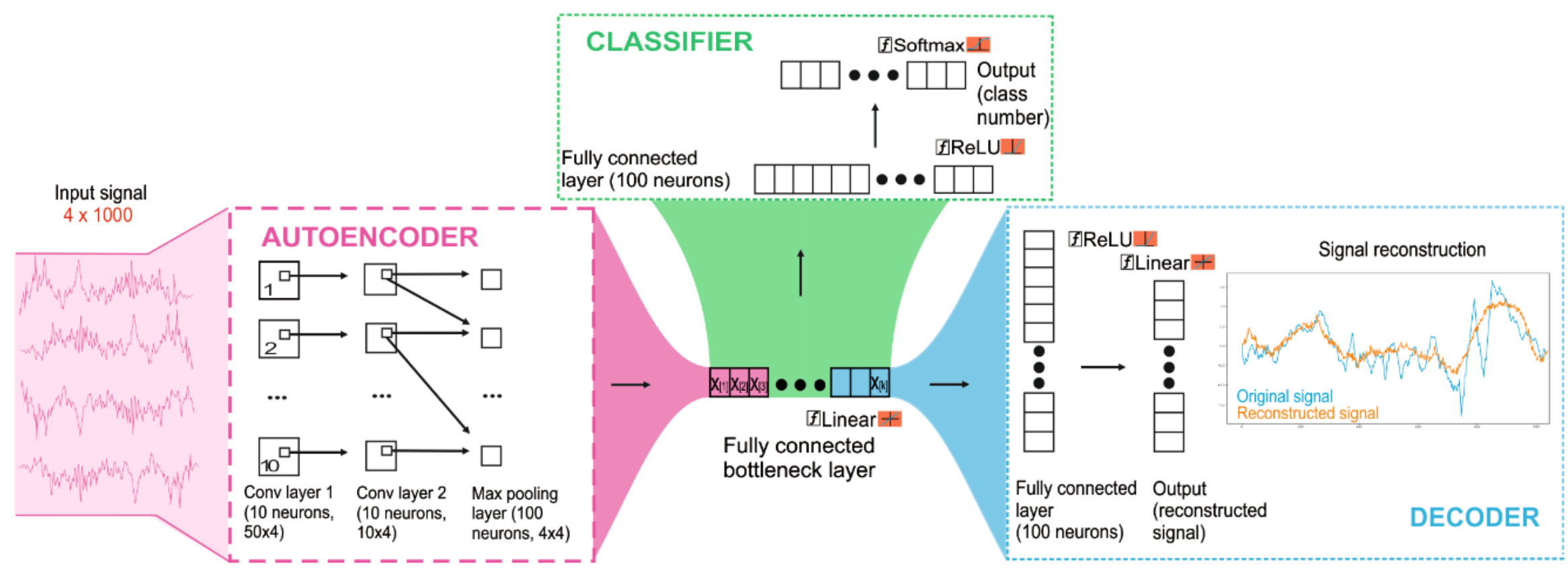

**Figure 4.** System architecture combining a classifier and an autoencoder. A special method of selective reduction of neurons allows the representation of electroencephalogram (EEG) signals in the bottleneck layer as a set of parameters. The classifier determines which features are more representative of pharmacological properties. Decoder reconstructs the original signal.

Four signals from different EEG channels were used as inputs in the network. These signals were converted to 1000 × 4 2D images and passed to the convolutional layer. The network was trained using the Adam method with a batch size of 100. There were two consecutive convolutional layers in the network. The first layer contained ten convolutional filters (neurons) with a kernel size of 50 × 4. The second convolutional layer also consisted of ten neurons with a kernel size of 20 × 4.

Let $y[n]$ be an input EEG signal. In the case of a four-channel signal, $y[n]$ has the form: $y[n] = (y_1[n], y_2[n], y_3[n], y_4[n])$, where the index at $y$ corresponds to the channel number, and $n = 0 ... 1000$ is the serial number of a discrete sampling point in time.

The signal $y[n]$ was appended with $L$ extra zeros. The new signal ŧ is denoted by $Y^{(0)}[n]$ and serves as input to the first convolutional layer. The output of the first convolutional layer is as follows:

$$z_r^{(1)}[n] = \mathrm{ReLU}\left( \sum_{l=0}^{L-1} w_r^{(1)}[l] Y^{(0)}[l+n] + b_r^{(1)} \right)$$

where $L$ is the length of the convolutional filters, $r = 0, 1, ..., R - 1$, ($R = 10$ is the number of convolutional filters), $w_i^{(1)}[l]$ is the impulse response of the filters, and $b_r^{(1)}$ represents bias.

The input signal in the second convolutional layer is $Y_r^{(1)}[m]$, where ($m = 0, 1, \ldots, 1000$) was obtained by appending $L$ zeros to $z_r^{(1)}[n]$:

$$z_r^{(2)}[m] = \mathrm{ReLU}\left( \sum_{l=0}^{L-1} \sum_{r'=0}^{R-1} w_{rr'}^{(2)}[l] Y_{r'}^{(1)}[l+m] + b_r^{(2)} \right)$$

where $w_{rr'}^{(2)}[m]$ is the impulse response of convolutional filters and $b_r^{(2)}$ represents biases. The second convolutional layer passes the outputs to the max-pooling layer, which reduces the size of the $z_r^{(2)}[m]$ signal 16 times at the expense of dropping out the smallest values. The output signal is then converted to a one-dimensional array $Y^{(2)}[q]$. This array is compressed by the fully connected encoder layer into a set of parameters $x[k]$:

$$x[k] = -b^{(code)}[k] - \sum_{q=0}^{Q-1} w_q^{(code)}[k] Y^{(2)}[q]$$

where $w_q^{(code)}$ and $b^{(code)}$ are the weights and biases, respectively, of the encoder layer. In this case, $k = 0 \ldots K$, where $K$ is the number of signal parameters. Q is the length of $Y^{(2)}[q]$.

Next, the signal $x[k]$ is transmitted to the decoder layer to reconstruct the original signal $y[n]$ by minimizing the error:

$$E_{autoencoder} = \sqrt{\sum_{n=0}^{N} \left(y'[n] - y[n]\right)^2},$$

In addition, the signal $x[k]$ is transmitted to the classifier layer to identify the drug class. Each EEG signal used in training has its one-hot number.

The output of the classification layer is denoted as $p[c]$, where $c = 0 ... P$, and P is the number of drugs.

Classification error during training was defined as the cross-entropy between the values $p[c]$ obtained by the network and the real class numbers $p^{true}[c]$, in which the signal $y[n]$ corresponds to:

$$E_{classifier} = H(p[c], p^{true}[c]).$$

During training on a set of signals recorded upon exposure to different drugs, the neural network clustered the signals in the multidimensional space, where the parameters $x[k]$ were coordinates in this space.

The center of the cluster corresponding to an arbitrary drug A was determined as follows:

$$x^A[k] = \frac{1}{M} \sum_{i=0}^{M} x^i[k]$$

where M is the number of signals for each drug in the dataset.

The distance between the clusters, i.e., the similarity of two arbitrary drugs A and B was determined using the following formula:

$$s_{AB} = \sum_{k=1}^{K} \left(x^A[k] - x^B[k]\right)^2$$

The closer the clusters were in the parameter space, the more similar the EEG signals.

The network learning cycle consisted of two stages. The autoencoder was trained first, i.e., the weights of the convolutional and fully connected layers were adjusted, the output of which was a reconstructed signal. At the second stage, the classifier was trained, i.e., the weights of convolutional layers and a fully connected layer were adjusted, the output of which was the drug number (class).

To determine the minimum number of parameters required to approximate and classify the EEG signal, as well as to interpret these parameters, we used a new method of selective reduction of neurons in the bottleneck layer.

The uniqueness of the proposed method relies on the type of dropout used during training, in which neurons were not turned off in a random order but are disabled from the end. The difference between the classic dropout and truncation for an arbitrary fully connected layer is illustrated in Figure 5.

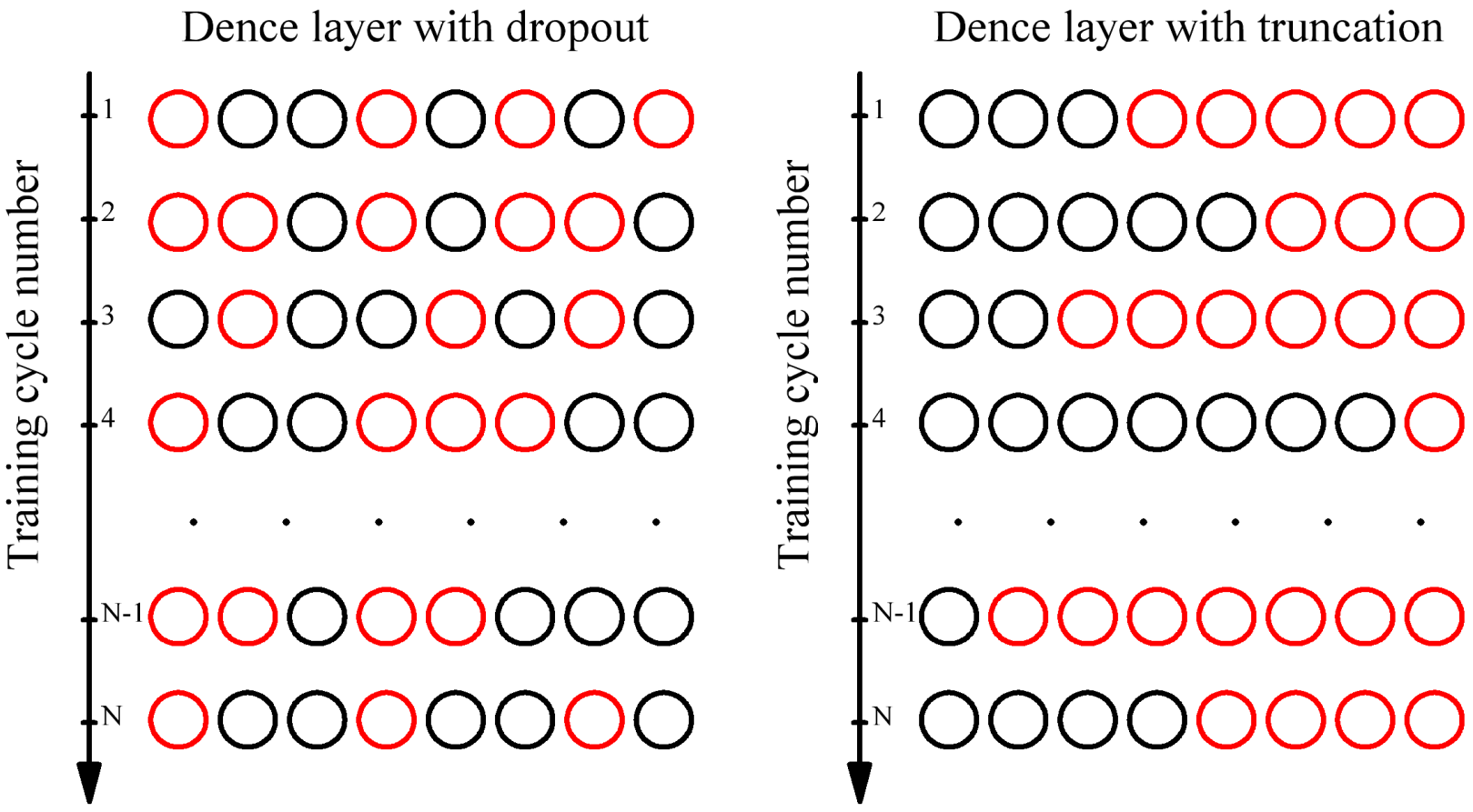

**Figure 5.** Dropout and truncation. Neurons are represented by circles. The black circles indicate active neurons while the red denotes neurons that were disabled during training.

This training method assumes that neurons in the bottleneck layer were selected according to their contribution to the results. Neurons with small $n$ values contributed the most, while those with high $n$ values contributes the least. By disabling neurons at the end of the layer, it was easy to plot the accuracy of signal reconstruction and the accuracy of classification depending on the size of the bottleneck, i.e., the number of parameters used to describe the EEG after just one full cycle of neural network training.

## 4 Results

### 4.1 Choosing the number of neurons in the bottleneck

To demonstrate the effectiveness of truncation and provide a rationale for the chosen architecture, we carried out several experiments with a simple training set consisting of signals for two drugs (arecoline and chloral hydrate). We determined the dependence of EEG signal reconstruction accuracy (defined as standard deviation) on the number of neurons ($n$) in the bottleneck with truncation and dropout. In both cases, this dependence was obtained by dropping out neurons from the end of the bottleneck (Figure 6). Further, we showed the relationship between the quality of the network operation and $n$ by manually setting the number of neurons in the bottleneck (started with one neuron and increased by ten up to 100 neurons) and re-training the network for each $n$ (brute force).

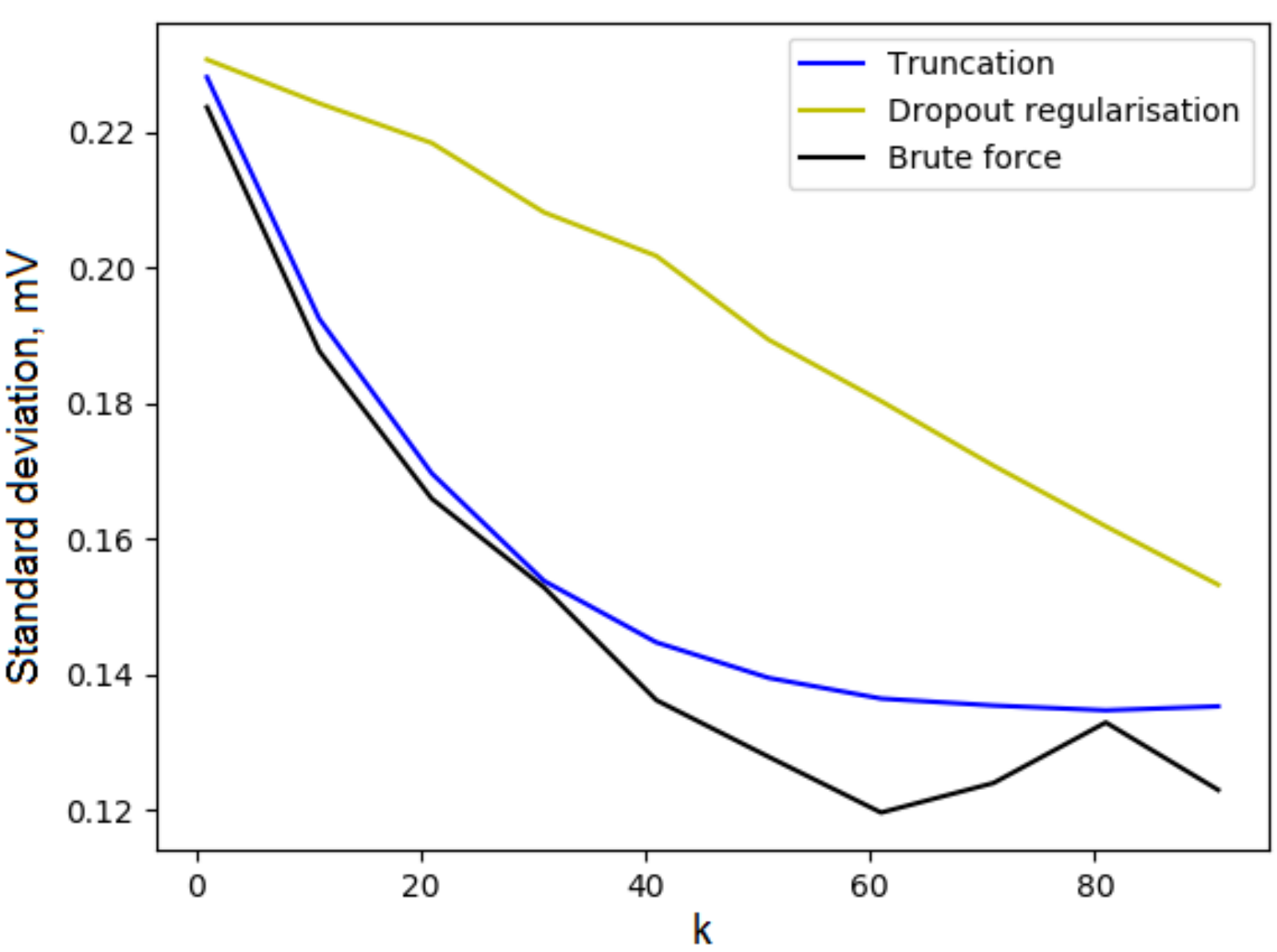

**Figure 6.** Dependence of the signal reconstruction accuracy on the number of neurons in the bottleneck layer. The curves for truncation and brute force

are almost the same, while the curve for truncation is built *n* times faster because in this case, the network needs to be trained only once. As can be seen in the graphs, the optimal choice is about 50–60 neurons in the bottleneck layer.

### 4.2 EEG classification

The classification accuracy was assessed using a 5-fold cross-validation method. The average probability of correct classification (for 11 classes) based on the cross-validation test was 0.58 ± 0.013 (mean ± standard deviation). The obtained classification accuracy significantly exceeded the accuracy for a random classifier (random guessing), which was 0.0909 for 11 drugs (11 classes).

The learning curves of the described network with truncation trained on the dataset that contained EEG signals from 11 drug groups are shown in Figure 7.

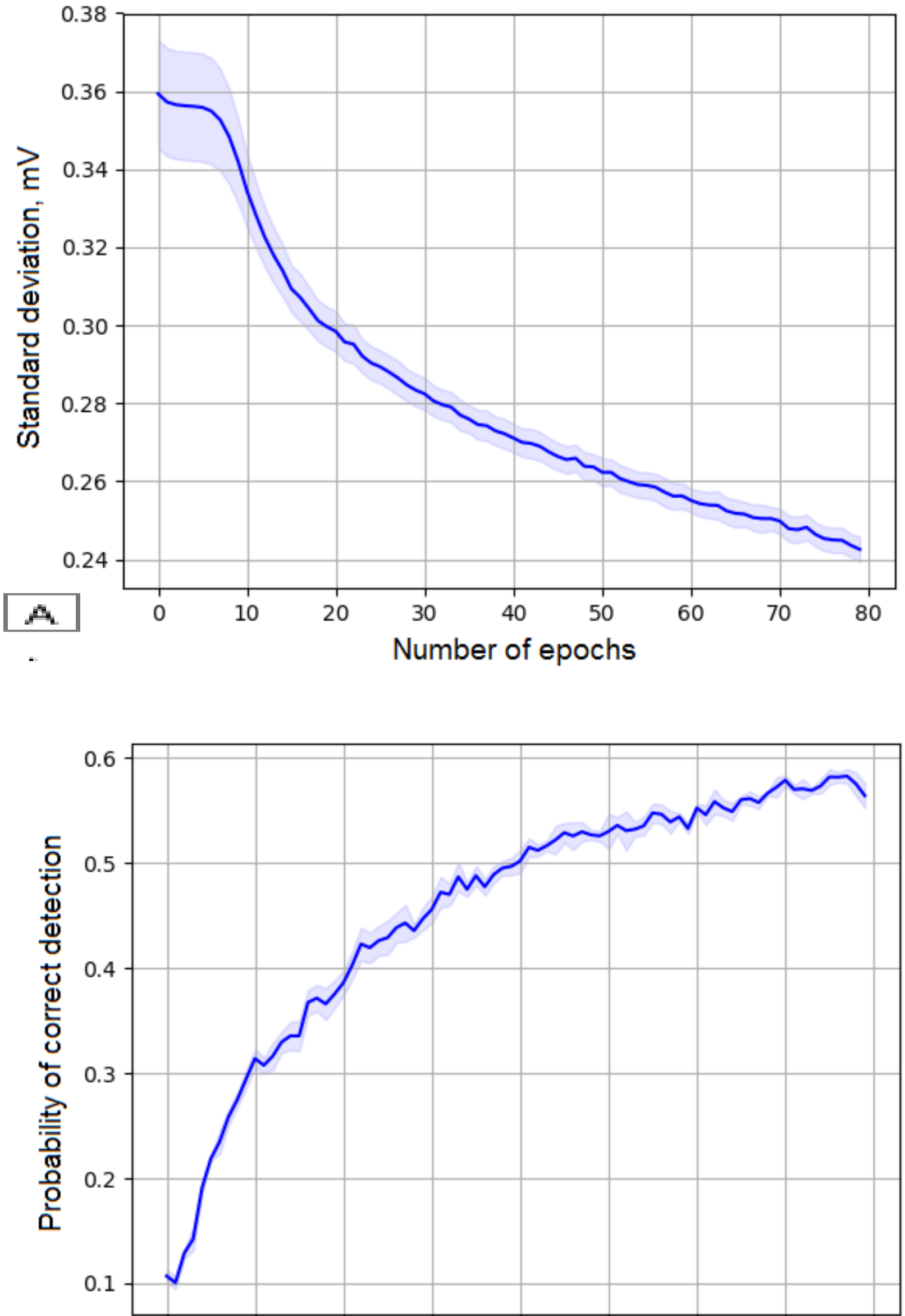

**Figure 7**. The learning curves of the neural network. (A) The signal reconstruction error (± standard deviation [SD]) of the test dataset during training. (B) Accuracy of the classifier (± SD) of the test dataset during training.

Using the described algorithm, we carried out experiments on the parameterization of the EEG signal with a layer of a fully connected autoencoder bottleneck. Using the neural network, we could represent a signal with a duration of up to 2 s as a function with 80 parameters (Figure 8). At the same time, the use of more than 20 parameters for classification was impractical.

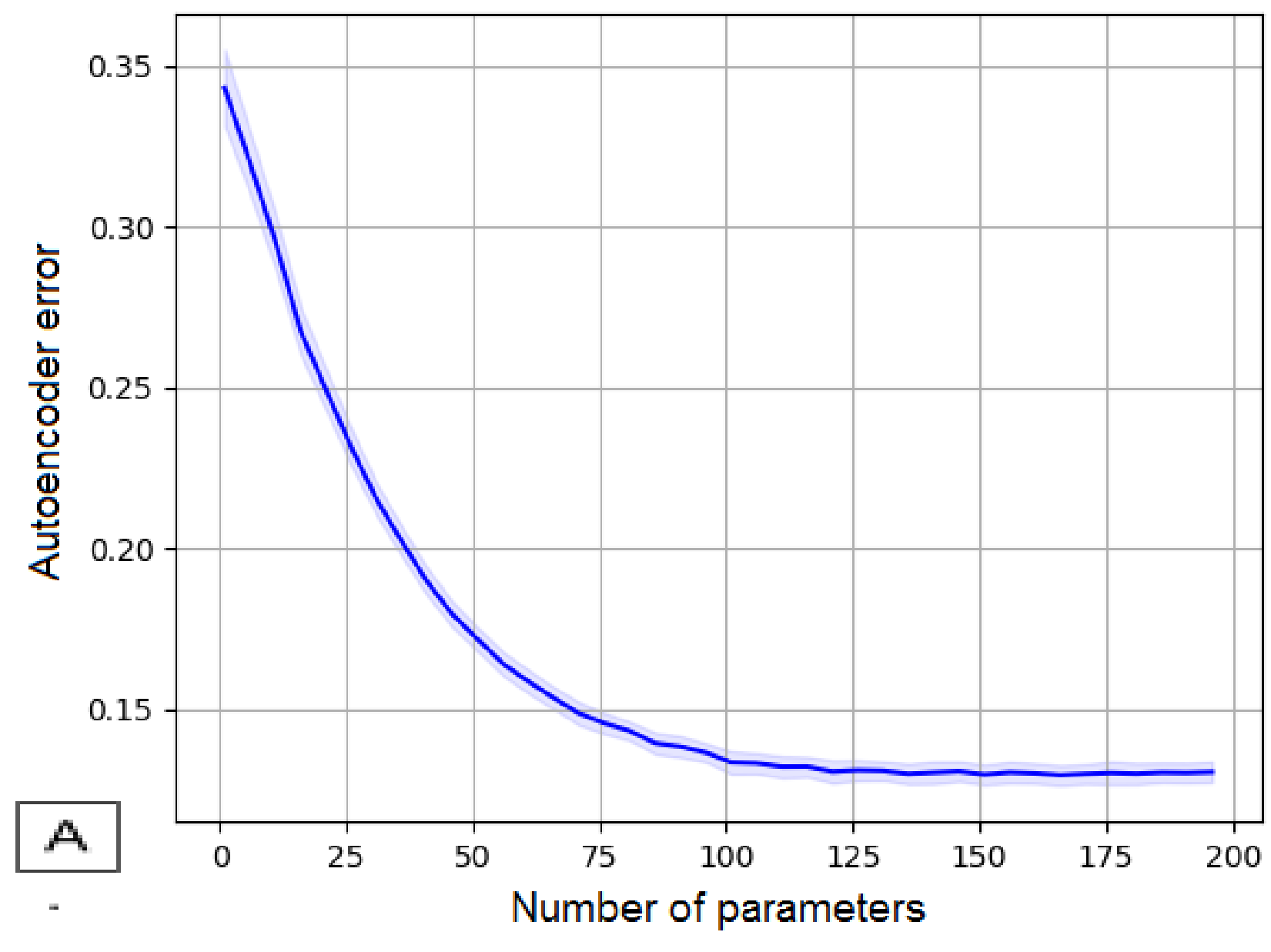

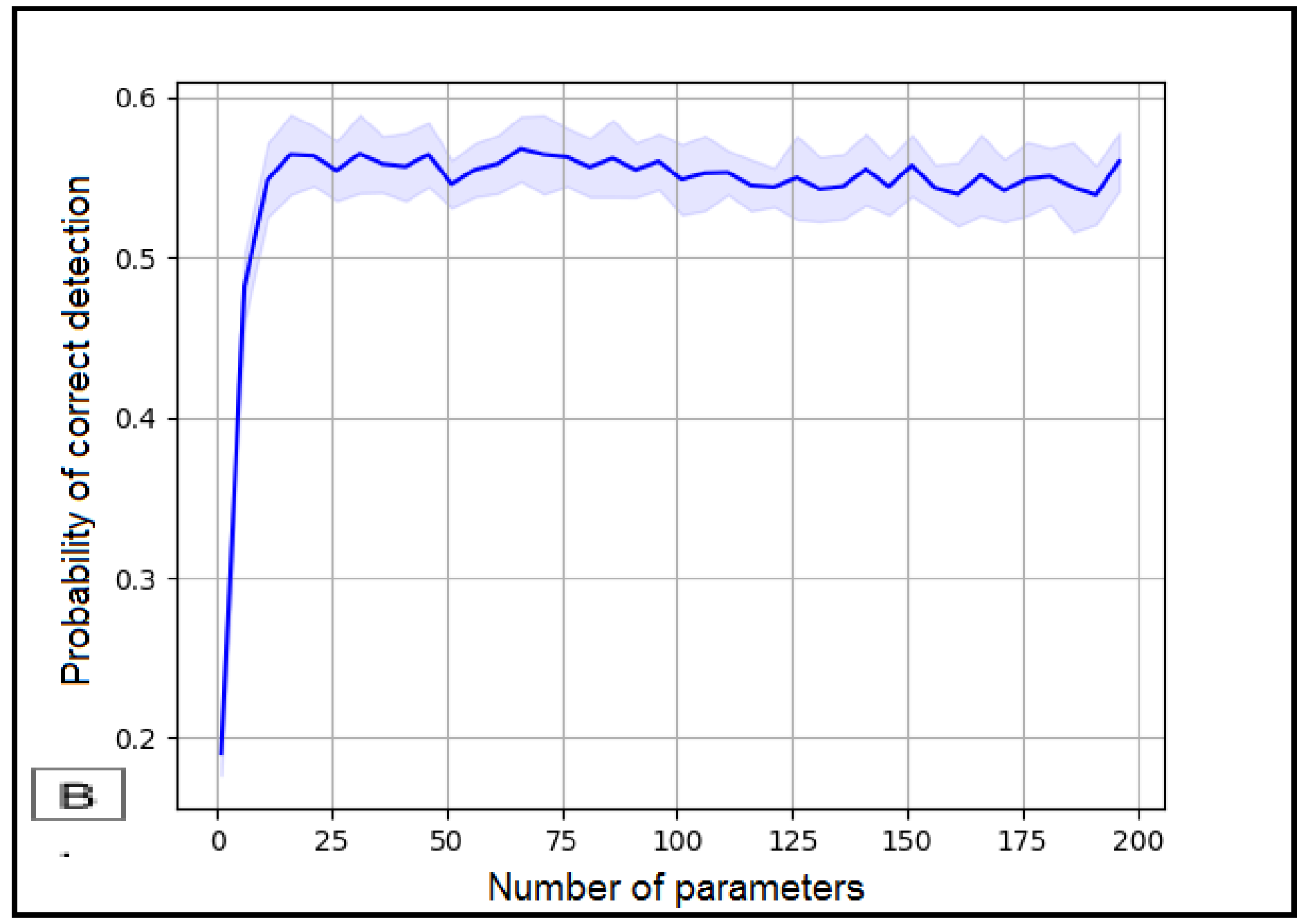

**Figure 8**. Determining the optimal number of parameters used by the network to approximate the electroencephalogram (EEG) signal. (A) Dependence of autoencoder error (± standard deviation [SD]) on the number of parameters in the bottleneck layer. (B) Dependence of classifier accuracy

(± SD) on the number of parameters. Values were calculated using the test dataset.

Using the network shown in Figure 5 only as a classifier for EEG signals recorded under drug exposure, it was possible to obtain a classification accuracy of at least 90% in the test set for four drugs with different mechanisms of action. The dependence of the classification accuracy of the test set signals on the number of drug classes is shown in Figure 9. The test set consisted of signals that were not used in the training set.

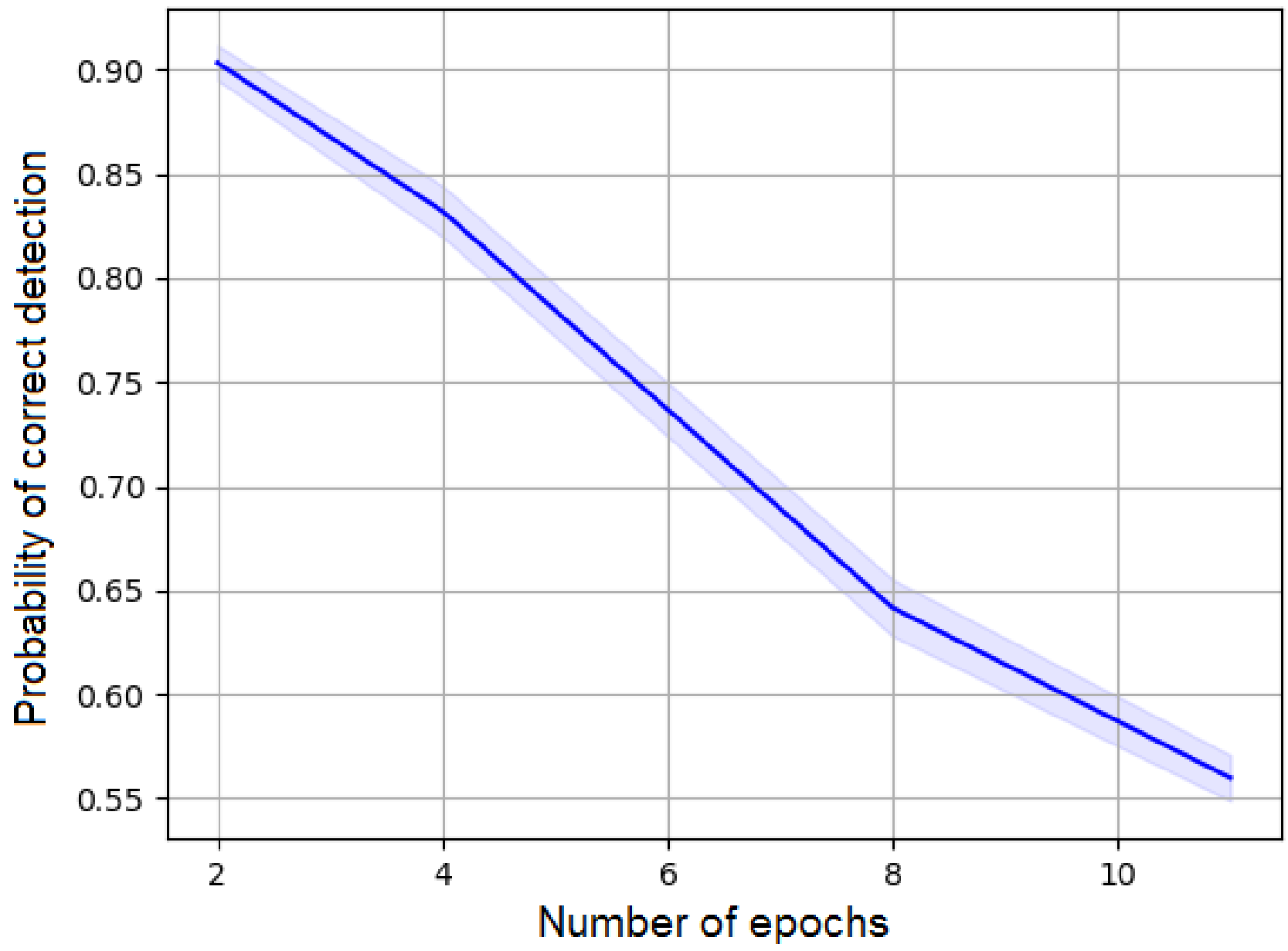

**Figure 9.** Dependence of the classification accuracy of drugs in the test dataset on the number of classified drugs (± standard deviation [SD]).

### 4.3 EEG clustering

The parameters obtained in the bottleneck layer are features of the EEG signal, which are descriptive of the pharmacological profile of a substance. Sample signals produced many parameter samples that approximately coincided, and are thus localized in clusters in the Euclidean space.

Parameter sets defined for drug classes in a 50-dimensional space (50 neurons in the bottleneck) were projected into 2D space using the t-SNE method [24]. The data structure is shown in Figure 10, where four distinct clusters corresponding to four drugs were visualized. The neural network architecture without a decoder tended to produce data that were visualized as more distinctly isolated clusters.

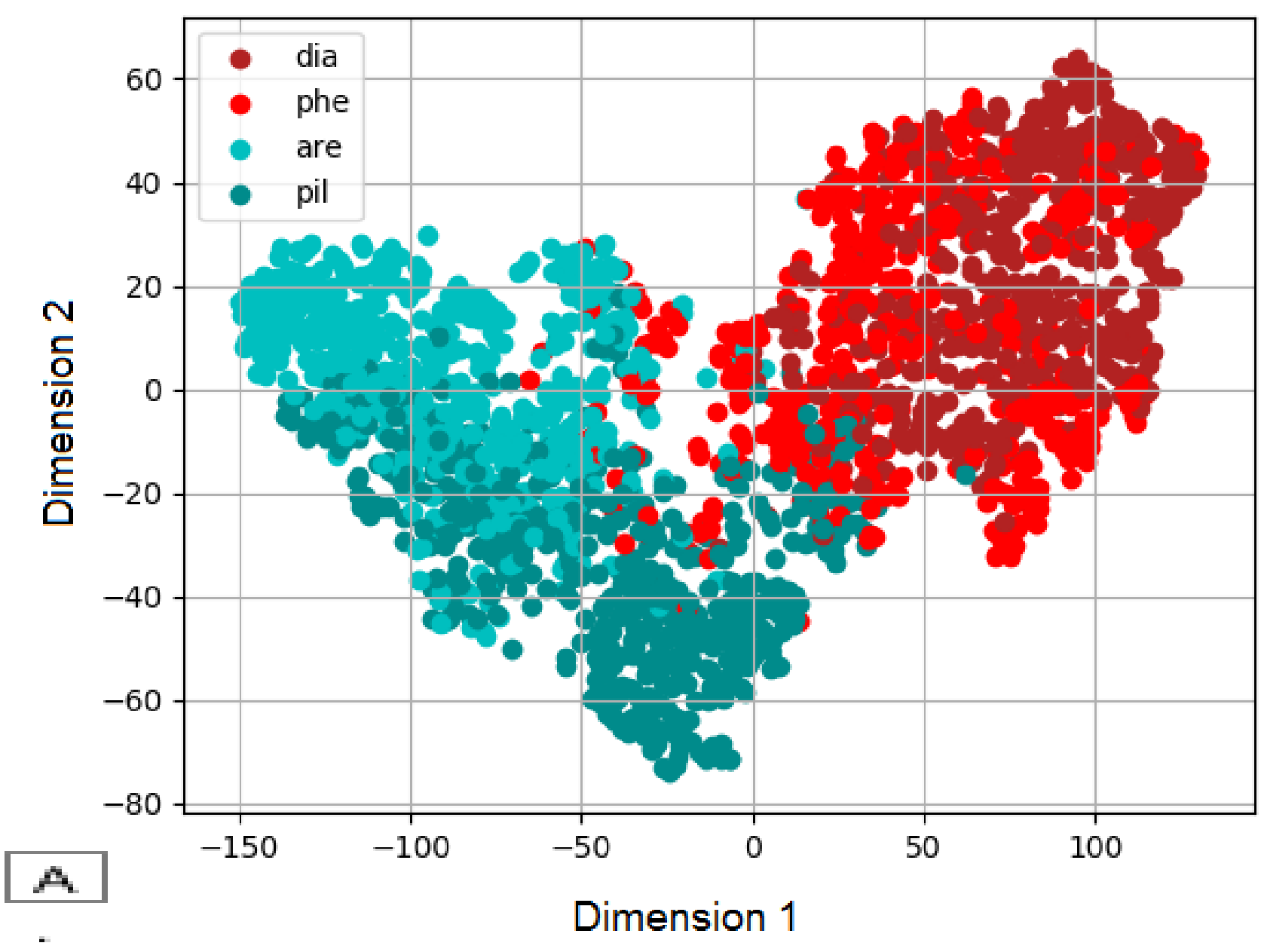

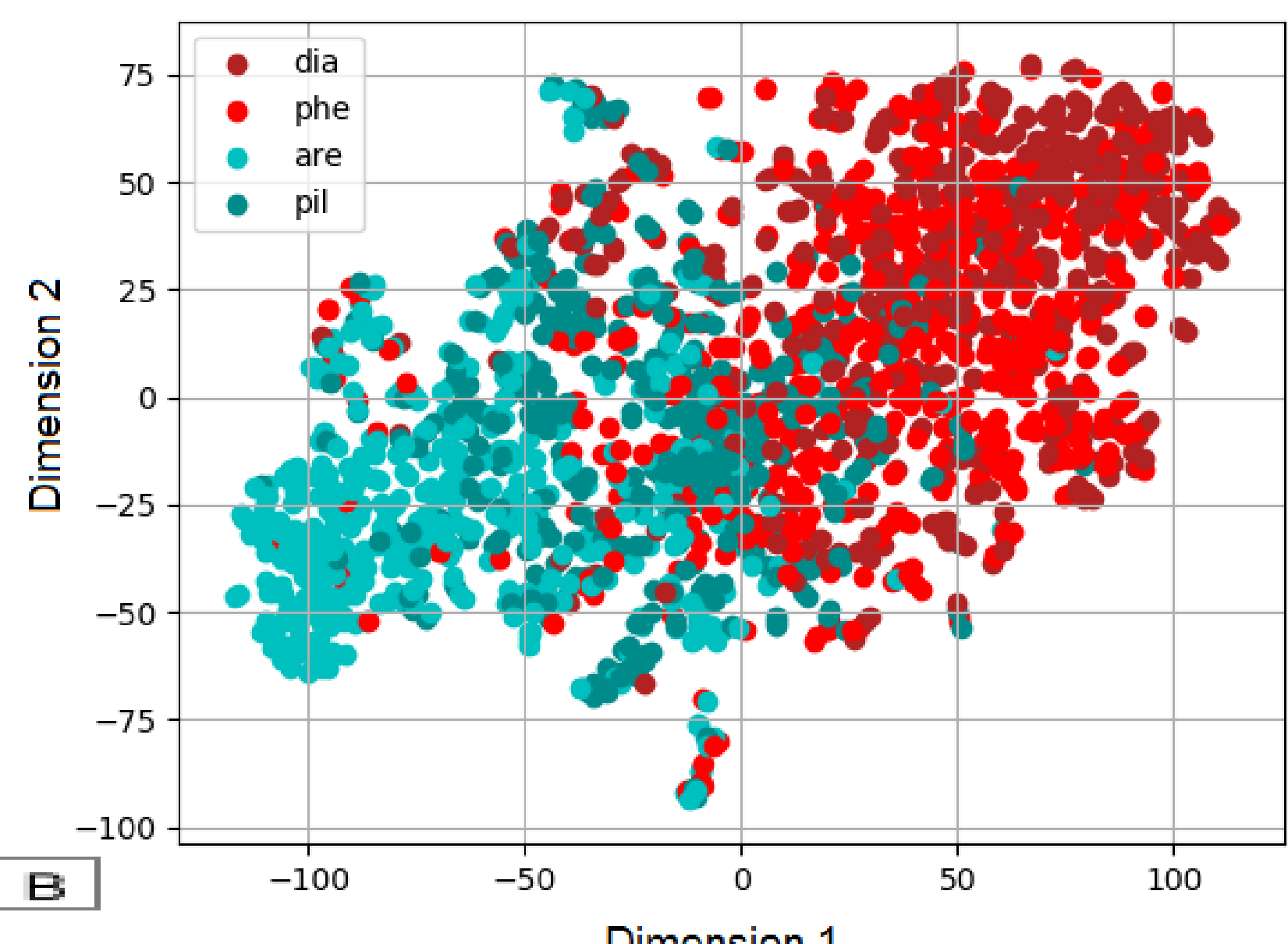

**Figure 10.** t-Distributed stochastic neighbor embedding (t-SNE) map of intracranial electroencephalogram (i-EEG) signal bottleneck parameters, with four highlighted drug classes. Plots generated with the perplexity of 5 and 3000 iterations. (A) Neural network architecture without a decoder. (B) Complete architecture (both a classifier and an autoencoder). dia, diazepam; phe, phenazepam; are, arecoline; pil, pilocarpine.

t-SNE visualization of a large dataset with 11 drug groups is shown in Figure 11. The cluster distribution patterns seemed less interpretable intuitively. However, it was observed that gamma-aminobutyric acid (GABA) receptor agonists and antagonists are located in opposite orientations from each other.

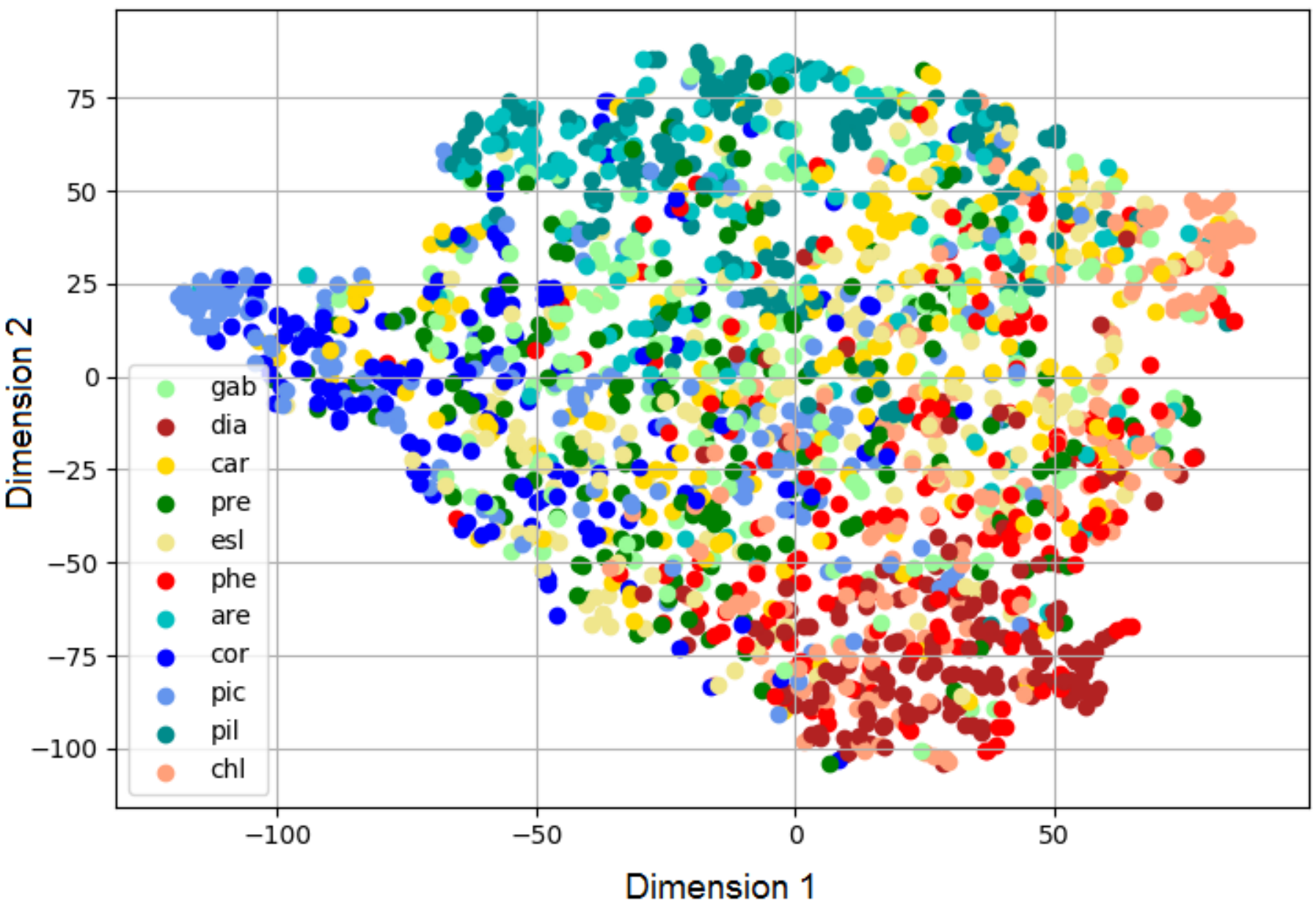

**Figure 11.** t-Distributed stochastic neighbor embedding (t-SNE) map of intracranial electroencephalogram (i-EEG) signal bottleneck parameters, with 11 highlighted drug classes. Plots generated with the perplexity of 5 and 3000 iterations. Network architecture without the decoder. gab, gabapentin; dia, diazepam; car, carbamazepine; pre, pregabalin; esl, eslicarbazepine; phe, phenazepam; are, arecoline; cor, corazol; pic, picrotoxin; pil, pilocarpine; chl, chloral hydrate.

Decreasing the dimension from 50 to 2 led to the loss of information. At the same time, the similarity of drugs could be defined as the distance between the clusters using the Euclidean distance (Figure 12).

The results confirmed that for a dataset with 11 drug classes, it was possible to correctly determine the qualitative distribution of the drug to the pharmacological group using the proposed clustering algorithm.

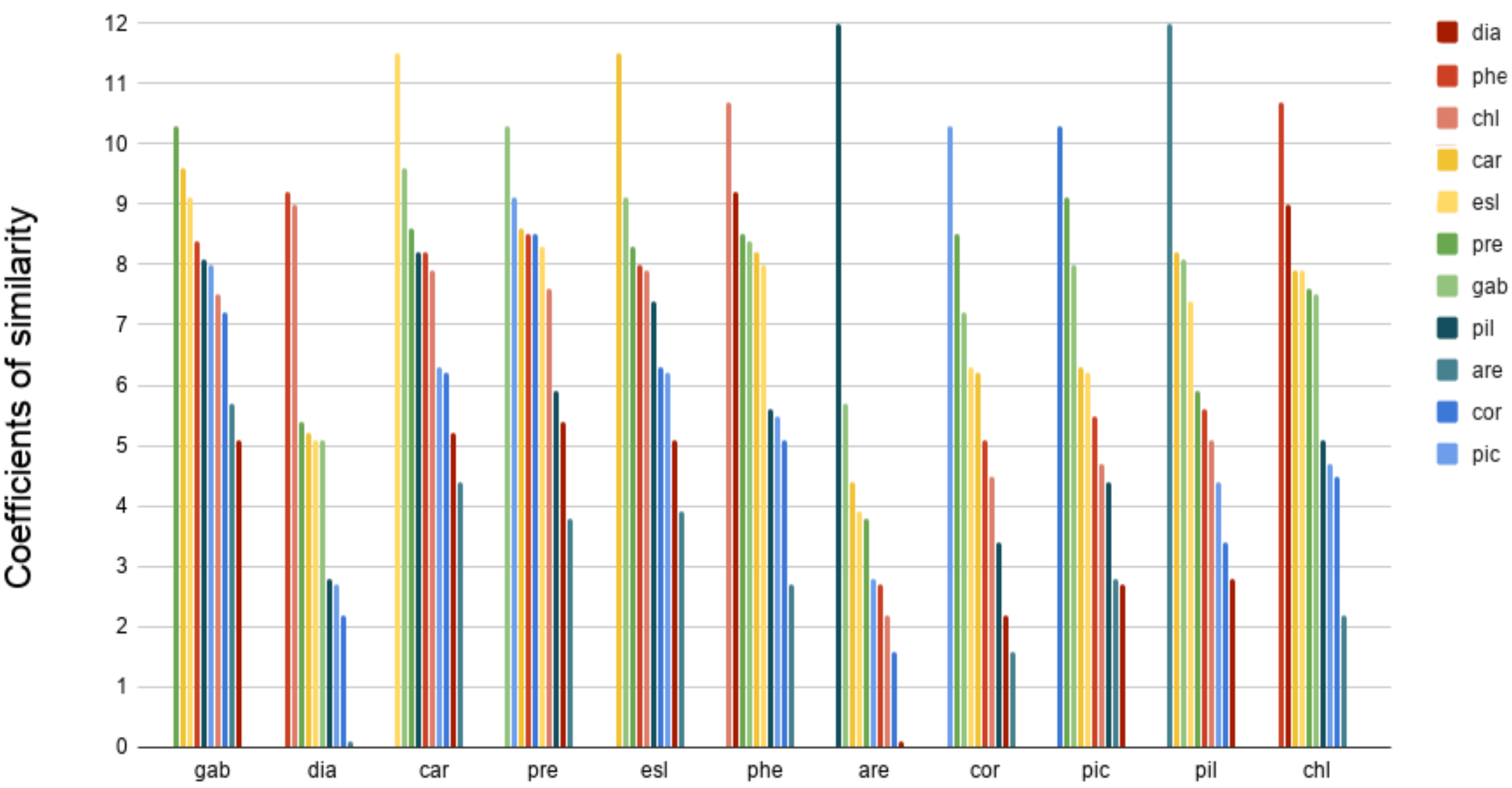

**Figure 12.** The relative proximity of clusters in the multidimensional parameter space as an indicator of the similarity of pharmacological properties of substances. The y-axis represents the coefficients of similarity of the drug effects on the electroencephalogram (EEG) signal. These coefficients are dimensionless, relative, and calculated as distances between clusters in unified multidimensional parameter space. The adjacent clusters were formed by drugs having a common pharmacological mechanism of action. dia, diazepam; phe, phenazepam; chl, chloral hydrate; car, carbamazepine; esl, eslicarbazepine; pre, pregabalin; gab, gabapentin; pil, pilocarpine; are, arecoline; cor, corazol; pic, picrotoxin.

Pregabalin was the most similar to gabapentin. Both drugs bind to an additional subunit ($\alpha 2\delta$) of voltage-gated calcium channels in the CNS, which are reportedly involved in epileptogenesis [28]. Eslicarbazepine and carbamazepine had an anticonvulsant effect, which was associated with inhibition of potential-dependent sodium channels in neurons of the brain.

EEG signals obtained after the administration of GABA-A receptor benzodiazepine site agonists (diazepam and phenazepam) were distributed close to a cluster of signals recorded upon chloral hydrate exposure, which is metabolized to the active component of trichloroethanol and implements its effects through the regulation of GABA-receptor complex [29]. Clusters of GABA-A agonists were at a maximum distance from GABA-A receptor antagonists, picrotoxin and corazole (pentylenetetrazole), as well as cholinomimetics, pilocarpine, and arecoline.

In general, the EEG signals obtained after the administration of drugs with an anticonvulsant effect had greater relative proximity to each other than those after the administration of drugs with a pro-convulsive effect.

## 5 Discussion

The neural network algorithm proposed here effectively recognized patterns of the neuronal response upon administration of various psychotropic drugs and could cluster substances based on the mechanism of action and similarities of their therapeutic effects.

i-EEG-mediated DTI prediction has an advantage in cases where classical DTI prediction approaches are ineffective; for example, in cases of multiple sclerosis, Parkinson's disease, and Alzheimer's disease, as well as other neurodegenerative and psychiatric diseases. This is because (a) an insufficient number of ligands are synthesized or not synthesized at all or (b) adequate targets are not identified.

In the present study, even in the absence of ligands with the same mechanism of action in the dataset, the algorithm could predict the possible effects of a substance, which opens up vast opportunities for the research and development of new anticonvulsants or other substances with psychotropic activity. In addition, there is a high potential for using this approach for repositioning drugs that are already used in clinical practice.

The doses of psychotropic substances were selected to exclude the development of severe impairment of consciousness and coordination of movements or convulsive phenomena that would lead to pronounced and easily recognizable signal changes. This allowed us to conclude that the proposed neural network has relatively high sensitivity and specificity, even under moderate pharmacological influence on brain function.

The productivity of this approach may be increased by replacing the animal subject with a nerve cell culture or living brain slice [30]. In future studies, the registration of neuronal activity may be carried out using *in vitro* microelectrode arrays (MEAs), such as planar MEAs and 3D HD-MEAs. The characteristics of the pharmacological response are not diminished or changed by the nature of dissociated neuronal networks when compared to *in vivo* models, i.e., MEAs can be used to study pharmacological effects on neural activity in a more controlled and straightforward environment [31].

The use of the decoder made it possible to extract features from the bottleneck that were needed not only for classification but also for obtaining the signal properties associated with the characteristics of research subjects and measuring equipment. If we used a decoder and the signals were still clustered (Figure 10), then we could say that the activation of neurons in the bottleneck reflected the slowly changing parameters of the EEG signal. Having determined the number of parameters describing the signal, we can estimate in subsequent studies the optimal requirements for a dataset, such as its combinatorial dimension, the number of individuals per drug, and the size of the signal sample for one drug. In addition, this approach made it possible to describe the EEG signal with a small number of parameters, which means that if we were to develop a methodology for interpreting these parameters, there would be new possibilities for EEG analysis.

## 6 Conclusion

We proposed an original approach for i-EEG-mediated DTI prediction using CNNs with a modified algorithm for selective reduction of neurons in the bottleneck layer. The suggested method associates a compound and a target or a compound and an effect through an intermediate link in the form of a bioelectric response to a substance.

The essence of this new approach lies in the fact that the nervous tissue is a perfect biological detector of psychotropic substances, as psychotropic action is directly reflected in bioelectric activity. The neural network algorithm efficiently recognizes patterns of neuronal response and can identify a substance via both the mechanism of action and therapeutic effect.

## Data availability and requirements

- Project name: ANN4EEG
- Project home page: https://cmi.to/ann4eeg/
- i-EEG data: http://dx.doi.org/10.17632/gmkbhj28jh.1 (274.7 Мб).
- Operating system (s): Platform independent.
- Programming language: Python.
- License: GNU GPL v3.

## Conflict of interest statement

The authors declare that they have no conflicts of interest.

## Acknowledgments

This research did not receive any specific grant from funding agencies in the public, commercial, or not-for-profit sectors.

The experimental protocols conformed to the European Convention for the Protection of Vertebrate Animals Used for Experimental and other Scientific Purposes (1986, ETS 123). Experiments were authorized by the local ethical committee.